\documentclass[10pt, twocolumn, twoside]{IEEEtran}

\usepackage{cite}
\usepackage{enumerate}
\usepackage{graphicx}
\usepackage[cmex10]{amsmath} 
\usepackage{amssymb}
\usepackage{mathtools}
\usepackage{xspace}
\usepackage{subfigure}
\usepackage[ruled,vlined]{algorithm2e} 
\usepackage{colonequals}
\usepackage[ mathscr]{euscript}
\usepackage{fixltx2e}    

\usepackage{epsfig}
\usepackage{epstopdf}

\usepackage{tikz}
\usepackage{pgfplots}

\usetikzlibrary{arrows,shapes,backgrounds,plotmarks,positioning}

\pgfplotsset{compat=newest}                         
\pgfplotsset{plot coordinates/math parser=false}
\newlength\figureheight
\newlength\figurewidth

\newtheorem{theorem}{Theorem}[section]
\newtheorem{lemma}[theorem]{Lemma}
\newtheorem{definition}[theorem]{Definition}
\newtheorem{corollary}[theorem]{Corollary}

\newtheorem{example}[theorem]{Example}
\newtheorem{remark}[theorem]{Remark}

\newcommand{\C}{\mathcal{K}}         
\newcommand{\Pak}{\mathcal{P}}       
\newcommand{\pv}{\mathbf{p}}         
\newcommand{\F}{\mathbb{F}}          
\newcommand{\Has}{\mathcal{H}}       
\newcommand{\Set}{\mathcal{S}}       
\newcommand{\rv}{\mathbf{r}}         
\newcommand{\R}{\mathcal{R}}         
\newcommand{\Rum}{\R_{\hat{\alpha}}}         
\newcommand{\Ru}{\R_{\alpha}}         
\newcommand{\Fu}{F_{\alpha}}         

\newcommand{\xv}{\mathbf{x}}         
\newcommand{\yv}{\mathbf{y}}         
\newcommand{\B}{\mathcal{B}}         
\newcommand{\splus}{\text{supp}^+}   
\newcommand{\sminus}{\text{supp}^-}  
\newcommand{\ev}{\mathbf{e}}         
\newcommand{\X}{\mathcal{X}}         
\newcommand{\Y}{\mathcal{Y}}         
\newcommand{\Zset}{\mathcal{Z}}
\newcommand{\Iset}{\mathcal{I}}

\newcommand{\ExAx}{\textbf{B-EXC\textsubscript{-}[Z]}}
\newcommand{\Dom}{\text{dom}_{\mathbf{z}}}

\newcommand{\N}{\mathbb{N}_0^K}
\newcommand{\Z}{\mathbb{Z}}
\newcommand{\Real}{\mathbb{R}}

\newcommand{\Zero}{\mathbf{0}}
\newenvironment{myindentpar}[1]%
 {\begin{list}{}%
         {\setlength{\leftmargin}{#1}}%
         \item[]%
 }
 {\end{list}}

\begin{document}

\title{Fairest Constant Sum-rate Transmission for Cooperative Data Exchange: An $M$-convex Minimization Approach}

\author{Ni~Ding, Rodney~A.~Kennedy and Parastoo~Sadeghi
\thanks{The authors are with the Research School of Engineering, College of Engineering and Computer Science, the Australian National University (ANU), Canberra, ACT 0200, Australia (email: $\{$ni.ding, rodney.kennedy, parastoo.sadeghi$\}$@anu.edu.au).}
}


\maketitle

\begin{abstract}
We consider the fairness in cooperative data exchange (CDE) problem among a set of wireless clients. In this system, each client initially obtains a subset of the packets. They exchange packets in order to reconstruct the entire packet set. We study the problem of how to find a transmission strategy that distributes the communication load most evenly in all strategies that have the same sum-rate (the total number of transmissions) and achieve universal recovery (the situation when all clients recover the packet set). We formulate this problem by a discrete minimization problem and prove its $M$-convexity. We show that our results can also be proved by the submodularity of the feasible region shown in previous works and are closely related to the resource allocation problems under submodular constraints. To solve this problem, we propose to use a steepest descent algorithm (SDA) based on $M$-convexity. By varying the number of clients and packets, we compare SDA with a deterministic algorithm (DA) based on submodularity in terms of convergence performance and complexity. The results show that for the problem of finding the fairest and minimum sum-rate strategy for the CDE problem SDA is more efficient than DA when the number of clients is up to five.
\end{abstract}


\section{Introduction}
Due to the growing amount of data exchange over wireless networks and increasing number of mobile clients, the base-station-to-peer (B2P) links are severely overloaded. It is called the `last mile' bottleneck problem in wireless transmissions. Cooperative peer-to-peer (P2P) communications is proposed for solving this problem. The idea is to allow mobile clients to exchange information with each other through P2P links instead of solely relying on the B2P transmissions. If the clients are geographically close to each other, the P2P transmissions could be more reliable than B2P ones.

Consider the situation when a base station wants to deliver a set of packets to a group of clients. Due to the fading effects of wireless channels, after several broadcasts via B2P links, there may still exist some clients that do not obtain all the packets. However, the clients' knowledge of the packet set may be complementary to each other. Therefore, instead of relying on retransmissions from the base station, the clients can broadcast combinations of the packets they know via P2P links so as to help the others recover the missing packets. For this kind of transmission method, there is a so-called cooperative data exchange (CDE) problem: how to find an efficient transmission strategy that achieves the \textit{universal recovery}, the situation when all clients recover the entire packet set.

Finding the minimum-sum rate strategy, the transmission scheme for universal recovery with the minimized total number of transmissions among clients, is the most commonly addressed CDE problem. It was introduced in \cite{Roua2010} and studied in \cite{SprintRand2010,AbediniNonMinRank2012,MiloDivConq2011}. There are also other optimization problems, e.g., finding the strategy that minimizes the weighted sum of transmission costs \cite{Ozgul2011,Court2011,Taj2011}. However, the solutions to most of these problems are usually not unique. On the other hand, in P2P communications, we wish to distribute the communication load as evenly as possible to prevent running out of one (or several) clients' battery usage and also promote incentives for the clients to cooperate in a fair manner. Therefore, a natural question that follows is how to find the fairest transmission strategy in a solution set.

In this paper, we study the problem of finding the fairest solution among the constant sum-rate strategy set, the set that contains all transmission schemes that achieve universal recovery and have the sum-rate equal to a constant number. We formulate this problem by a discrete minimization problem and prove its $M$-convexity: The feasible region is an $M$-convex set, and the objective function is $M$-convex. We show that the solution can be searched by a steepest descent algorithm (SDA) based on the optimality criterion of $M$-convex functions. We analyze the differences in convergence performance and complexity between SDA and a deterministic algorithm (DA) proposed in \cite{Milo2012} based on the submodularity. By applying both SDA and DA to the problem of finding the fairest solution in the minimum sum-rate strategies, we show that SDA converges faster and involves less complexity than DA when the number of clients is up to five.

\subsection{Related Works}

The fairness in CDE problem has also been studied in \cite{Taj2011,Milo2012}. In \cite{Taj2011}, the best solution in terms of Jain's fairness index is found by solving a convex minimization problem among those strategies that minimize the weighted sum of the transmission costs of clients. The CDE system in \cite{Taj2011} allows packet splitting, i.e., \cite{Taj2011} considers an optimization problem over continuous variables (real numbers). On the contrary, \cite{Milo2012} studies the fairness when packet splitting is not allowed, where a discrete minimization problem is formulated, and the submodularity of this problem is proved and used to propose a DA algorithm. This paper studies the same problem as in \cite{Milo2012}. The difference is that we prove the discrete convexity ($M$-convexity) of it, based on which an SDA algorithm is presented. In Section~\ref{sec:relation}, we show that the $M$-convexity can also be proved by the results derived in \cite{Milo2012} by the one-to-one correspondence between $M$-convex sets and submodular functions. In Section~\ref{sec:SDA}, we show the performance of SDA by comparing it to DA by examples.

\section{System Model and Problem Statement}
\label{sec:system}

\begin{figure}[tpb]
	\centering
    \scalebox{0.95}{\begin{tikzpicture}

\draw (-2.6,0.3) rectangle (-1.4,-0.3);
\node at (-2,0) {client $1$};
\draw (-2,0.3)--(-2,1)--(-1.7,1)--(-2,0.8)--(-2.3,1)--(-2,1);
\node at (-2,-0.5) {\scriptsize $\{\pv_1,\pv_2,\pv_3,\pv_4,\pv_5\}$};

\draw (2.6,0.3) rectangle (1.4,-0.3);
\node at (2,0) {client $2$};
\draw (2,0.3)--(2,1)--(1.7,1)--(2,0.8)--(2.3,1)--(2,1);
\node at (2,-0.5) {\scriptsize $\{\pv_1,\pv_2,\pv_6\}$};

\draw (-0.6,1.6) rectangle (0.6,1);
\node at (0,1.3){client $3$};
\draw (0,1.6)--(0,2.3)--(0.3,2.3)--(0,2.1)--(-0.3,2.3)--(0,2.3);
\node at (0,0.8) {\scriptsize $\{\pv_3,\pv_4,\pv_6\}$};

\node at (0,2.5) {$\phantom{a}$};

\end{tikzpicture}}
	\caption{An example of CDE system: There are three clients that want to obtain six packets. The has-sets are $\Has_1= \{\pv_1,\pv_2,\pv_3,\pv_4,\pv_5\}$, $\Has_2= \{\pv_1,\pv_2,\pv_6\}$ and $\Has_3=\{\pv_3,\pv_4,\pv_6\}$.}
	\label{fig:CDESystem}
\end{figure}
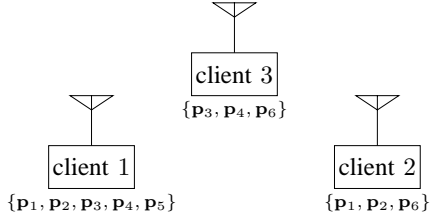

Let $\Pak=\{\pv_1,\dotsc,\pv_N\}$ be the packet set containing $N$ linearly independent packets. Each packet $\pv_i$ belongs to a field $\F_q$. The system contains $K$ geographically close clients. Define the client set as $\C=\{1,\dotsc,K\}$. Each client $j\in\C$ initially obtains $\Has_j\subset\Pak$. Here, $\Has_j$ is called the \textit{has-set} of client $j$. We also denote $\Has_j^{c}=\Pak\setminus\Has_j$ as the packet set that is missing at client $j$. The clients are assumed to collectively know the the packet set, i.e., $\cup_{j\in\C}\Has_j=\Pak$. The P2P links between clients are error-free, i.e., any information broadcast by client $j$ can be heard losslessly by client $j'$ for all $j'$ such that $j'\in\C\setminus{\{j\}}$. The clients broadcast linear combinations of the packets in their has-sets in order to help each other to recover the entire packet set $\Pak$. For example, in the CDE system in Fig.~\ref{fig:CDESystem},\footnote{The CDE system in Fig.~\ref{fig:CDESystem} is also the example considered in \cite{Milo2012}. We use this system because we will compare our work with the results in \cite{Milo2012} in Section~\ref{sec:relation} and \ref{sec:SDA}.} client $1$ broadcasting $\pv_1+\pv_3$ helps client $2$ recover $\pv_3$ and client $3$ recover $\pv_1$, and client $2$ broadcasting $\pv_1+\pv_6$ helps client $1$ recover $\pv_6$ and client $3$ recover $\pv_1$.

Let $\rv=(r_1,\dotsc,r_K)$ be a transmission strategy, where $r_j$ denotes the total number of linear combinations transmitted by client $j$. We call $\sum_{j\in\C}r_j$ the sum-rate of strategy $\rv$. Let $\hat{\alpha}$ be the minimum sum-rate that allows universal recovery. Denote $\alpha$ as an integer constant and assume that $\alpha\geq\hat{\alpha}$. Consider the CDE problem when the sum-rate is constrained to a budget $\alpha$, i.e., the problem of finding a transmission strategy that achieves universal recovery and has a sum-rate equal to constant $\alpha$. The solution to this problem is not unique in general. For example, the CDE system in Fig.~\ref{fig:CDESystem} has $\hat{\alpha}=4$. Let $\alpha=\hat{\alpha}=4$ and consider two transmission schemes: one is that client $1$ broadcasts $\pv_1+\pv_3$, $\pv_2+\pv_4$ and $\pv_5$ and client $3$ broadcasts $\pv_6$; The other is that client $1$ broadcasts $\pv_2+\pv_4$ and $\pv_5$, client $2$ broadcasts $\pv_1+\pv_6$ and client $3$ broadcast $\pv_3$. The transmission strategies associated with the two schemes are $(3,0,1)$ and $(2,1,1)$, respectively. Both strategies achieves universal recovery and have sum-rate equal to $4$. However, strategy $(3,0,1)$ is not as good as $(2,1,1)$ in terms of fairness: In strategy $(3,0,1)$, the energy consumption at client $1$ is high while client $2$ is a free-rider. So, there arises a problem of how to find a fairest solution in the constant sum-rate transmission strategy set.

\section{Discrete Convex Minimization}


\subsection{Discrete Minimization Problem}

It is proved in \cite{Court2010,Court2010M} that the necessary and sufficient condition for a transmission strategy $\rv$ to achieve universal recovery is that $\sum_{j\in\Set}r_j\geq|\bigcap_{j\in\C\setminus\Set}\Has_j^{c}|$ for all $\Set$ such that $\Set\subset\C$. Denote
\begin{equation}
\rv(\Set)=\sum_{j\in\Set}r_j.
\end{equation}
Let $\Ru$ be the set that contains all strategies that allow universal recovery and have sum-rate equal to $\alpha$. We can describe $\Ru$ as
\begin{multline}
\Ru=\Big\{\rv\in\N \colon \rv(\Set)\geq \Big| \bigcap_{j\in\C\setminus\Set}\Has_j^{c} \Big|, \forall\Set\subset\C,\\
        \sum_{j\in\C}r_j=\alpha \Big\}.
\end{multline}
Note, for all $\alpha\geq\hat{\alpha}$, set $\Ru$ is nonempty, and when $\alpha=\hat{\alpha}$, the problem under consideration is to find the fairest solution in the minimum sum-rate strategy set $\Rum$.

Let $\Fu\colon\Ru\mapsto\Real$ be the fairness measurement functions. We assume that $\Fu$ is a \textit{discrete separable convex} function in $\Ru$, i.e., $\Fu$ is in the form of
\begin{equation}
\Fu(\rv)= \begin{cases}
            \sum_{j\in\C}f_j(r_j) & \rv\in\Ru  \\
            +\infty                   & \rv\notin\Ru
        \end{cases},
\end{equation}
where $f_j$ is convex in $r_j$ for all $j$. In TABLE~\ref{tab:Fairs}, we show some examples of $f_j$ based on different types of fairness indices. In this paper, we use the uniform fairness definition $f_j(r_j)=r_j\log r_j$.\footnote{Although we just consider $f_j(r_j)=r_j\log r_j$, it should be clear that the results derived in the paper is applicable to other definitions in TABLE~\ref{tab:Fairs}.} The fairest strategy that achieves universal recovery and has sum-rate equal to $\alpha$ can be searched by solving the minimization problem
\begin{equation} \label{eq:obj}
\min_{\rv} \Fu(\rv).
\end{equation}
Denote $\Dom{f}=\{\xv\in\Z^K \colon -\infty<f(\xv)<+\infty\}$ the \textit{effective domain} of function $f$. $\Dom{\Fu}=\Ru$, i.e., $\Dom{\Fu}$ denotes the feasible region of \eqref{eq:obj}. 

\begin{table}[tbp]
	\renewcommand{\arraystretch}{1.3}
	\caption{examples of definitions of $f_j$}
	\label{tab:Fairs}
	\centering
	\begin{tabular}{c c}
	\hline\hline
	$f_j(r_j)$ & \textbf{type of fairness} \\ %
	\hline
	$r_j^2/\alpha^2$ & Jain's fairness \cite{Jain1984}  \\
    $-\log r_j$ & proportional fairness \cite{Kushner2004}\\
    $r_j\log r_j$ & uniform fairness \cite{Milo2012}\\ [0.3ex]
    \hline
	\end{tabular}
\end{table}

\subsection{$M$-convexity}
\label{sec:Main}
In this section , we show the $M$-convexity of \eqref{eq:obj}. We first clarify some definitions as follows.

\begin{definition}[$M$-convex set\cite{Murota2003}] \label{def:Mconset}
A set $\B\subseteq\Z^K$ is $M$-convex if it satisfies the following exchange axiom:
\begin{myindentpar}{0.5cm}
\ExAx: For all $\xv,\yv\in\B$ and $u\in\splus(\xv-\yv)$, there exists $v\in\sminus(\xv-\yv)$ such that $\xv-\ev_u+\ev_v\in\B$.
\end{myindentpar}
\end{definition}
In \ExAx, $\splus(\xv)$ and $\sminus(\xv)$ are the \textit{positive and negative supports} of $\xv$, respectively, and defined as
\begin{align}
\splus(\xv)=\{ j \colon x_j>0, j\in\C  \},  \nonumber \\
\sminus(\xv)=\{ j \colon x_j<0, j\in\C \},  \nonumber
\end{align}
where $x_j$ is the $j$th entry of $\xv$. $\ev_j$ is the unit vector with all entries being $0$ except the $j$th entry being $1$.

\begin{definition}[$M$-convex function\cite{Murota2003}]  \label{def:Mconfun}
A function $f\colon \Z^K\mapsto{\Real_+}$ is $M$-convex if $\Dom{f}\neq\emptyset$ is an $M$-convex set and for all $\xv,\yv\in\Dom{f}$ and $u\in\splus(\xv-\yv)$ there exists $v\in\sminus(\xv-\yv)$ such that
\begin{equation} \label{eq:MconInEq}
f(\xv)+f(\yv) \geq f(\xv-\ev_u+\ev_v)+f(\yv+\ev_u-\ev_v).
\end{equation}
\end{definition}

\begin{remark}
$M$-convex set is a class of discrete convex sets based on \ExAx. Let $\overline{\B}\subseteq{\Real^K}$ be the convex hull of $\B$. If $\B$ is an $M$-convex set, $\B=\overline{\B}\cap\Z^K$, i.e., all integer points contained in $\overline{\B}$ constitutes $\B$. Alternatively speaking, $\B$ describes a hole-free region in $\Z^K$ \cite{Murota2003}. $M$-convex function is a class of discrete convex functions defined based on \ExAx. We will show examples of $M$-convex sets and functions in Example~\ref{ex:MConSet}. 
\end{remark}

In the following context, we prove the $M$-convexity of \eqref{eq:obj} by showing the $M$-convexity of $\Ru$ and $\Fu$. We start the proof by showing a property of the tight set as follows.

We call $\Set\subset\C$ a \textit{tight set} of $\rv$ if $\rv(\Set)=\Big| \bigcap_{j\in\C\setminus\Set}\Has_j^{c} \Big|$. A tight set has the following property.
\begin{lemma} \label{lemma:Tset}
Let $\rv\in\Ru$ and $\X,\Y\subset\C$ such that $\X\cap\Y\neq\emptyset$ and $\X\cup\Y\neq\C$. if $\X$ and $\Y$ are tight sets, then $\X\cap\Y$ is a tight set.
\end{lemma}
\begin{IEEEproof}
Since $\rv(\X)+\rv(\Y)=\rv(\X\cap\Y)+\rv(\X\cup\Y)$, we have
\begin{align}  \label{eq:lemma}
\rv(\X\cap\Y)&=\rv(\X)+\rv(\Y)-\rv(\X\cup\Y)    \nonumber\\
           &\leq \Big| \bigcap_{j\in\C\setminus\X}\Has_j^{c} \Big| + \Big| \bigcap_{j\in\C\setminus\Y}\Has_j^{c} \Big| - \Big| \bigcap_{j\in\C\setminus\X\cup\Y}\Has_j^{c} \Big|   \nonumber \\
           &\leq \Big| \bigcap_{j\in\C\setminus\X\cap\Y}\Has_j^{c} \Big|.
\end{align}
But, $\rv(\X\cap\Y)\geq \Big| \bigcap_{j\in\C\setminus\X\cap\Y}\Has_j^{c} \Big|$ since $\rv\in\Ru$. Therefore, $\rv(\X\cap\Y)=\Big| \bigcap_{j\in\C\setminus\X\cap\Y}\Has_j^{c} \Big|$. Note, the last inequality in \eqref{eq:lemma} is because of Corollary~\ref{coro:app} in Appendix~\ref{app:lemma}.
\end{IEEEproof}

\begin{theorem} \label{theo:Mconset}
$\Ru$ is an $M$-convex set.
\end{theorem}
\begin{IEEEproof}
We use an approach similar to the proof of Proposition 4.14 in \cite{Murota2003}. Assume that the exchange axiom \ExAx\ does not hold in $\Ru$, i.e., for some $\xv,\yv\in\Ru$, there exists $u\in\splus(\xv-\yv)$ such that $\xv-\ev_u+\ev_v\notin\Ru$ for all $v\in\sminus(\xv-\yv)$. Since the sum-rate of $\xv-\ev_u+\ev_v$ equals $\alpha$ always, the only situation that could make $\xv-\ev_u+\ev_v\notin\Ru$ is that for each $v\in\sminus(\xv-\yv)$, there exist $\Set_v\subset\C$ such that $u\in\Set_v$, $v\notin\Set_v$ and
\begin{align}
\xv-\ev_u+\ev_v(\Set_v)=\xv(\Set_v)-1 \leq \Big| \bigcap_{j\in\C\setminus\Set_v}\Has_j^{c} \Big|.
\end{align}
But, $\xv(\Set_v)\geq\Big| \bigcap_{j\in\C\setminus\Set_v}\Has_j^{c} \Big|$. Therefore, $\xv(\Set_v)=\Big| \bigcap_{j\in\C\setminus\Set_v}\Has_j^{c} \Big|$, i.e., $\Set_v$ is a tight set for all $v\in\sminus(\xv-\yv)$. Consider the set $\Zset=\bigcap_{v\in\sminus(\xv-\yv)}\Set_v$. Since $\Set_v\cap\Set_{v'}\neq\emptyset$ and $\Set_v\cup\Set_{v'}\neq\C$ for all $v,v'\in\sminus(\xv-\yv)$. By Lemma~\ref{lemma:Tset}, $\Zset$ is a tight set, i.e., $\xv(\Zset)=\Big| \bigcap_{j\in\C\setminus\Zset}\Has_j^{c} \Big|$. Also, since $u\in\Zset$ and $v\notin\Zset$ for all $v\in\sminus(\xv-\yv)$,
\begin{equation} \label{eq:app}
\yv(\Zset)<\xv(\Zset)=\Big| \bigcap_{j\in\C\setminus\Zset}\Has_j^{c} \Big|.
\end{equation}
\eqref{eq:app} contradicts the condition that $\yv\in\Ru$. Therefore, for all $u\in\splus(\xv-\yv)$, there must exist $v\in\sminus(\xv-\yv)$ such that $\xv-\ev_u+\ev_v\in\Ru$, i.e., \ExAx\ is satisfied in $\Ru$. By Definition~\ref{def:Mconset}, $\Ru$ is an $M$-convex set.
\end{IEEEproof}

\begin{theorem} \label{theo:Mconfun}
$\Fu(\rv)$ is an $M$-convex function.
\end{theorem}
\begin{IEEEproof}
Since each discrete separable convex function is also $M$-convex \cite{Murota2003}, $\Fu(\rv)$ is an $M$-convex function. Also, $\Dom{\Fu}=\Ru$ is nonempty and $M$-convex. By Definition~\ref{def:Mconfun}, $\Fu(\rv)$ is an $M$-convex function.
\end{IEEEproof}

\begin{corollary}
\eqref{eq:obj} is an $M$-convex minimization problem.
\end{corollary}
\begin{IEEEproof}
This is a direct result of Theorem~\ref{theo:Mconfun}.
\end{IEEEproof}

\begin{remark} \label{rem:Mconfun}
Similar to continuous convex minimization problems, for $M$-convex minimization problems, local optimality guarantees global optimality, which is based on the \textit{optimality criterion}\cite{Murota2003}
\begin{equation} \label{eq:OptCri}
\Fu(\rv)\leq\Fu(\rv-\ev_u+\ev_v), \forall\rv\in\Dom\Fu.
\end{equation}
\end{remark}

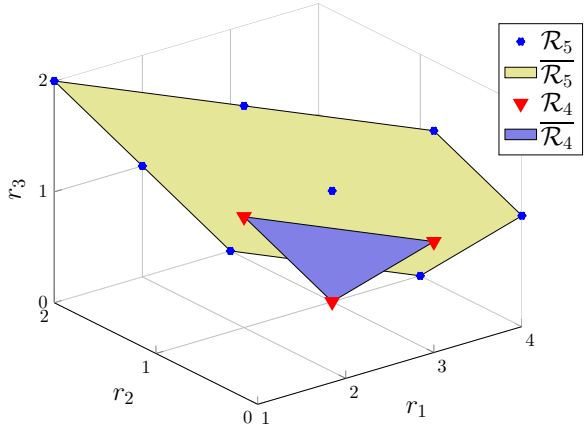
\begin{figure}[tbp]
	\centering
    \scalebox{0.7}{
%
%
%
\definecolor{mycolor1}{rgb}{0.5,0.5,0.9}%
\definecolor{mycolor2}{rgb}{0.9,0.9,0.6}%
\begin{tikzpicture}

\begin{axis}[%
width=3.5in,
height=3in,
view={322.5}{30},
scale only axis,
xmin=1,
xmax=4,
xlabel={\Large $r_1$},
xmajorgrids,
ymin=0,
ymax=2,
ylabel={\Large $r_2$},
ytick={0, 1, 2},
ymajorgrids,
zmin=0,
zmax=2,
zlabel={\Large $r_3$},
ztick={0, 1, 2},
zmajorgrids,
axis x line*=bottom,
axis y line*=left,
axis z line*=left,
legend style={at={(0.95,0.95)},anchor=north west,draw=black,fill=white,legend cell align=left}
]

\addplot3 [
color=blue,
line width=3.0pt,
only marks,
mark=asterisk,
mark options={solid}]
table[row sep=crcr] {
1 2 2\\
2 1 2\\
2 2 1\\
3 0 2\\
3 1 1\\
3 2 0\\
4 0 1\\
4 1 0\\
};
\addlegendentry{\Large $\R_5$};

\addplot3[area legend,solid,fill=mycolor2,draw=black]
table[row sep=crcr]{
x y z\\
1 2 2 \\
2 1 2 \\
3 0 2 \\
4 0 1 \\
4 1 0 \\
3 2 0 \\
2 2 1 \\
1 2 2\\
};
\addlegendentry{\Large $\overline{\R_5}$};

\addplot3 [
color=red,
line width=3.0pt,
only marks,
mark=triangle,
mark options={solid,,rotate=180}]
table[row sep=crcr] {
2 1 1\\
3 0 1\\
3 1 0\\
};
\addlegendentry{\Large $\R_4$};

\addplot3[area legend,solid,fill=mycolor1,draw=black]
table[row sep=crcr]{
x y z\\
2 1 1 \\
3 0 1 \\
3 1 0 \\
2 1 1 \\
};
\addlegendentry{\Large $\overline{\R_4}$};

\end{axis}
\end{tikzpicture}%
}
	\caption{$\R_4$ and $\R_5$ of the CDE problem in Fig.~\ref{fig:CDESystem}. $\overline{\R_4}$ and $\overline{\R_5}$ are the convex hull of $\R_4$ and $\R_5$, respectively. It can be seen that $\R_4=\overline{\R_4}\cap\Z^K$ and $\R_5=\overline{\R_5}\cap\Z^K$.}
	\label{fig:SampRateRegion}
\end{figure}

\begin{example} \label{ex:MConSet}
Consider the CDE system in Fig.~\ref{fig:CDESystem} when $\alpha=4$. We have
\begin{equation}
\R_4= \Big\{ (2,1,1),(3,0,1),(3,1,0) \Big\}.
\end{equation}
It can be shown that $\R_4$ satisfies \ExAx. For example, consider $\xv=(2,1,1)$ and $\yv=(3,0,1)$. $\splus(\xv-\yv)=\{2\}$ and $\sminus(\xv-\yv)=\{1\}$. $\xv-\ev_2+\ev_1=(3,0,1)\in\R_4$. It can be checked that this property applies to all $\xv,\yv\in\R_4$. When $\alpha=5$, we have
\begin{multline}
\R_5=\Big\{ (1,2,2),(2,1,2),(2,2,1),(3,0,2),  \\
     (3,1,1),(3,2,0),(4,0,1),(4,1,0) \Big\},
\end{multline}
where \ExAx\ also holds. In Fig.~\ref{fig:SampRateRegion}, we show the set $\R_4$, $\R_5$ and their convex hulls $\overline{\R_4}$ and $\overline{\R_5}$. It can be seen that $\overline{\R_4}$ and $\overline{\R_5}$ lie on the planes $\sum_{j\in\C}r_j=4$ and $\sum_{j\in\C}r_j=5$, respectively, and $\R_4$ and $\R_5$ are hole-free, i.e., all the integer points lie on $\overline{\R_4}$ and $\overline{\R_5}$ belong to $\R_4$ and $\R_5$, respectively. Consider the function $F_4(\rv)$ in Example~\ref{ex:MConSet}. We have
\begin{equation}
F_4(\rv)=\begin{cases}
            1.3963 & \rv=(2,1,1)  \\
            3.2958 & \rv=(3,0,1) \text{ or } \rv=(3,1,0)  \\
            +\infty & \text{otherwise}
         \end{cases}.
\end{equation}
Therefore, $\arg\min_{\rv}F_4(\rv)=\{(2,1,1)\}$. Similarly, we can show that $\arg\min_{\rv}F_5(\rv)=\{(1,2,2),(2,1,2),(2,2,1)\}$. It can be verified that $F_4$ and $F_5$ satisfy the optimality criterion in Remark~\ref{rem:Mconfun}.
\end{example}

\section{Relationship with Existing Works}
\label{sec:relation}
In \cite{Milo2012,Court2011}, the properties of the set $\Ru$ has been studied. They both show that $\Ru$ is related to a submodular set function. In this section, we show that the Theorem~\ref{theo:Mconset} can be proved by the results in \cite{Milo2012,Court2011} and \eqref{eq:obj} is in fact a constrained resource allocation problem.

\subsection{$M$-convex set and submodularity}
\label{sec:Rset}

We first clarify the associated definitions as follows.

\begin{definition}[submodular set function \cite{Fujishige2005}]
Let $2^{\C}$ be the power set (the set of all subsets) of $\C$, $f\colon 2^{\C}\mapsto \Real_+$ is submodular if for all $\X,\Y\subseteq\C$
\begin{equation} \label{eq:SubMInEq}
f(\X)+f(\Y) \geq f(\X\cap\Y)+f(\X\cup\Y).
\end{equation}
\end{definition}

\begin{definition}[polyhedron and base polyhedron \cite{Fujishige2005}]
For a function $f\colon 2^{\C}\mapsto \Real_+$, the polyhedron $P(f)$ and base polyhedron $B(f)$ are defined as
\begin{align}
P(f)&=\{\tilde{\rv}\in\Real_+^{K} \colon \tilde{\rv}(\Set) \leq f(\Set),\Set\subseteq\C \},  \nonumber \\
B(f)&=\{\tilde{\rv}\in P(f) \colon \tilde{\rv}(\C)=f(\C) \}.    \nonumber
\end{align}
If $f$ is submodular, $P(f)$ and $B(f)$ are submodular polyhedron and submodualr base polyhedron, respectively.
\end{definition}

In \cite{Milo2012,Court2011}, it was shown that $\Ru=B(\grave{g}_\alpha)\cap\Z^K$, where 
\begin{equation}
\grave{g}_\alpha(\Set)=\begin{cases}
                            0                                                          & \Set=\emptyset \\
                            \alpha- \Big| \bigcap_{j\in\C\setminus\Set} \Has_j^c \Big| & \text{otherwise}
                       \end{cases}     
\end{equation}
was a crossing submodular function.\footnote{A crossing submodular function $\grave{f}$ satisfies the inequality \eqref{eq:SubMInEq} for all $\X,\Y\subseteq\C$ such that $\X\cap\Y\neq\emptyset$, $\X-\Y\neq\emptyset$, $\Y-\X\neq\emptyset$ and $\X\cup\Y\neq\C$.} Let
\begin{multline} \label{eq:gSubM}
g_\alpha(\Set)=\min \Big\{ \sum_{i\in \Iset}\acute{g}_\alpha(\Y_i) \colon \{\Y_i:i\in\Iset=\{1,2,\dotsc\}\} \\
         \text{ is a partition of } \Set \Big\}.
\end{multline}
According to Theorem 2.6 in \cite{Fujishige2005}, $B(g_\alpha)=B(\grave{g}_\alpha)$, and $g$ is submodular. Alternatively speaking, $\Ru=B(g_\alpha)\cap\Z^K$, i.e., $\Ru$ can be fully determined by submodular set function $g_\alpha$. In the following context, we discuss the relationship between the submodularity of $g_\alpha$ and the $M$-convexity of $\Ru$.

\begin{theorem}[one-to-one correspondence \cite{Murota2003}] \label{theo:Relation}
The class of submodular set functions $f$ and the class of $M$-convex sets $\B$ are in one-to-one correspondence through the mutually inverse mappings:
\begin{align}
f\mapsto\B \colon & \B=B(f)\cap\Z^K,   \nonumber \\
\B\mapsto f \colon & f(\Set)=\sup\{\rv(\Set) \colon \rv\in\B \},\forall{\Set\subseteq\C}.  \nonumber
\end{align}
\end{theorem}
In Theorem~\ref{theo:Relation}, the base polyhedron $B(f)$ is exactly the convex hull of $\B$\cite{Murota2003}. Therefore, the interpretation of mapping $f\mapsto\B$ is: For every submodular function, its base polyhedron determines the convex hull of an $M$-convex set. Based on Theorem~\ref{theo:Relation}, Theorem~\ref{theo:Mconset} in Section~\ref{sec:Main} can also be proved by the results in \cite{Milo2012,Court2011}. Since $\Ru=B(g_\alpha)\cap\Z^K$, due to the submodularity of $g_\alpha$, $\R_\alpha$ is $M$-convex.
%

\begin{figure}[tbp]
	\centering
    \scalebox{0.7}{
%
%
%
\definecolor{mycolor1}{rgb}{0.5,0.5,0.9}%
\begin{tikzpicture}

\begin{axis}[%
width=3.5in,
height=3in,
view={322.5}{30},
scale only axis,
xmin=0,
xmax=3.5,
xlabel={\Large $r_1$},
xtick={0, 1, 2},
xmajorgrids,
ymin=0,
ymax=1.5,
ylabel={\Large $r_2$},
ytick={0, 1, 2},
ymajorgrids,
zmin=0,
zmax=1.5,
zlabel={\Large $r_3$},
ztick={0, 1, 2},
zmajorgrids,
axis x line*=bottom,
axis y line*=left,
axis z line*=left,
legend style={at={(0.95,0.88)},anchor=north west,draw=black,fill=white,legend cell align=left}
]
\addplot3 [
color=red,
line width=3.0pt,
only marks,
mark=triangle,
mark options={solid,,rotate=180}]
table[row sep=crcr] {
2 1 1\\
3 0 1\\
3 1 0\\
};
\addlegendentry{\Large $\R_4$};

\addplot3[area legend,solid,fill=mycolor1,draw=black]
table[row sep=crcr]{
x y z\\
2 1 1 \\
3 0 1 \\
3 1 0 \\
2 1 1 \\
};
\addlegendentry{\Large $B(g_4)$};

\addplot3[area legend,solid,fill=white!90!black,opacity=4.000000e-01,draw=black]
table[row sep=crcr]{
x y z\\
0 0 0 \\
3 0 0 \\
3 1 0 \\
0 1 0 \\
0 0 0 \\
};
\addlegendentry{\Large $P(g_4)$};

\addplot3[solid,fill=white!90!black,opacity=4.000000e-01,draw=black,forget plot]
table[row sep=crcr]{
x y z\\
0 0 0 \\
0 1 0 \\
0 1 1 \\
0 0 1 \\
0 0 0 \\
};

\addplot3[solid,fill=white!90!black,opacity=4.000000e-01,draw=black,forget plot]
table[row sep=crcr]{
x y z\\
0 0 0 \\
3 0 0 \\
3 0 1 \\
0 0 1 \\
0 0 0 \\
};

\addplot3[solid,fill=white!90!black,opacity=4.000000e-01,draw=black,forget plot]
table[row sep=crcr]{
x y z\\
0 0 0 \\
3 0 0 \\
3 0 1 \\
0 0 1 \\
0 0 0 \\
};

\addplot3[solid,fill=white!90!black,opacity=4.000000e-01,draw=black,forget plot]
table[row sep=crcr]{
x y z\\
3 0 0 \\
3 1 0 \\
3 0 1 \\
3 0 0 \\
};

\addplot3[solid,fill=white!90!black,opacity=4.000000e-01,draw=black,forget plot]
table[row sep=crcr]{
x y z\\
0 1 0 \\
0 1 1 \\
2 1 1 \\
3 1 0 \\
0 1 0 \\
};

\addplot3[solid,fill=white!90!black,opacity=5.000000e-01,draw=black,forget plot]
table[row sep=crcr]{
x y z\\
0 0 1 \\
3 0 1 \\
2 1 1 \\
0 1 1 \\
0 0 1 \\
};

\end{axis}
\end{tikzpicture}%
}
	\caption{The submodular polyhedron $P(g_4)$ and submodular base polyhedron $B(g_4)$ for the CDE problem in Fig.~\ref{fig:CDESystem}. $g_4$ is given in \eqref{eq:gSubMSamp}. It can be seen that $B(g_4)$ is exactly $\overline{\R_4}$, the convex hull of $\R_4$ and $\R_4$ contains all the integer points in $B(g_4)$, i.e., $\R_4=B(g_4)\cap\Z^K$. }
	\label{fig:SubMPolyH}
\end{figure}
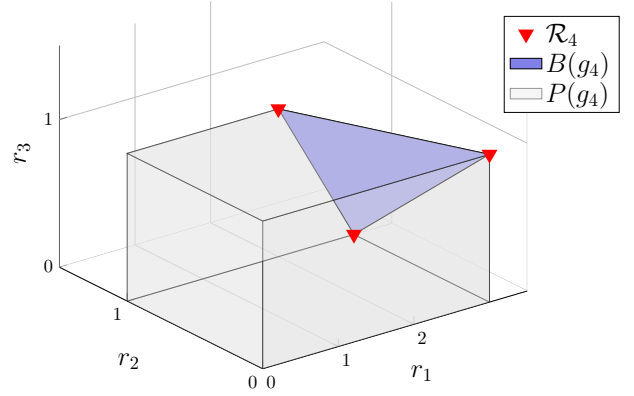

\begin{example}
In the CDE system in Fig.~\ref{fig:CDESystem}. The function $g_\alpha$ determined by \eqref{eq:gSubM} when $\alpha=4$ is \cite{Milo2012}
\begin{align} \label{eq:gSubMSamp}
&g_4(\emptyset)=0,g_4(\{1\})=3,g_4(\{2\})=1,g_4(\{3\})=1, \nonumber \\
&g_4(\{1,2\})=4,g_4(\{1,3\})=4,g_4(\{2,3\})=2,   \nonumber \\
&g_4(\{1,2,3\})=4.
\end{align}
We show the polyhedron $P(g_4)$ and base polyhedron $B(g_4)$ in Fig.~\ref{fig:SubMPolyH}. It can be seen that $B(g_4)$ is the intersection between $P(g_4)$ and plane $\sum_{j\in\C}\tilde{r}_j=4$. By comparing Fig.~\ref{fig:SubMPolyH} to Fig.~\ref{fig:SampRateRegion}, it can be seen that $B(g_4)$ is in fact $\overline{\R_4}$, the convex hull of the $M$-convex set $\R_4$, and $\R_4=B(g_4)\cap\Z^K$.
\end{example}

In fact, this paper and \cite{Milo2012,Court2011} present two different ways of proving the $M$-convexity of $\R_\alpha$: We prove the $M$-convexity of $\R_\alpha$ by definition; The authors in \cite{Milo2012,Court2011} express $\R_\alpha$ by a submodular base polyhedron, which implicitly proves the $M$-convexity of $\R_\alpha$ by Theorem~\ref{theo:Relation}.

\subsection{$M$-convexity of Resource Allocation Problems}
\label{sec:Rfunc}
Since the concepts of $M$-convex set and submodular base polyhedron are exchangeable based on the one-to-one correspondence in Theorem~\ref{theo:Relation}, the results in Section~\ref{sec:Main} also implicitly indicate the relationship between problem~\eqref{eq:obj} and resource allocation problems under submodular constrtains. Based on the results in Section~\ref{sec:Rset}, we can rewrite problem \eqref{eq:obj} as
\begin{equation} \label{eq:ResAll}
\min \Big\{ F_\alpha(\rv) \colon \rv\in B(g_\alpha) \Big\}.
\end{equation}
This is called the \textit{sparable convex resource allocation problem under submodular constrains}\cite{Ibaraki1988} (since the feasible region can be described by a submodular base polyhedron). The $M$-convexity of this problem has been proved in \cite{Murota1998,Murota2003}. In \cite{Shioura2004,Tamura2005,Moriguchi2011}, various algorithms based on $M$-convexity have been developed for solving problem~\eqref{eq:ResAll}.

\section{Steepest Descent Algorithm}
\label{sec:SDA}

One of the advantages of solving an $M$-convexity minimization problem is that efficient methods can be derived. This section considers one of the simplest methods, the steepest descent algorithm (SDA). We analyze the convergence performance and time complexity of SDA by comparing it to the deterministic algorithm (DA) proposed in \cite{Milo2012}.

	\begin{algorithm} [t]
	\label{algo:SDA}
	\small
	\SetAlgoLined
	\SetKwRepeat{Repeat}{repeat}{until}
    \SetKw{Return}{return}
	\BlankLine
        $k=0$\;
        Find a $\rv^{(0)}\in\Ru$\;
        \Repeat{$\Fu(\rv^{(k+1)})\geq\Fu(\rv^{(k)})$}{
            Find the descent direction $-\ev_{u^*}+\ev_{v^*}$ by determining $(u^*,v^*)$ such that
            \begin{align}
                (u^*,v^*)\in&\arg\min\{\Fu(\rv^{(k)}-\ev_u+\ev_v) \colon \nonumber \\
                            &u,v\in\C, u\neq{v}, \rv^{(k)}+\ev_u-\ev_v\in\Ru \};  \nonumber
            \end{align}
            $\rv^{(k+1)}=\rv^{(k)}-\ev_{u^*}+\ev_{v^*}$\;
            }
        \Return minimizer $\rv^*=\rv^{(k)}$\;
	
	\caption{Steepest Descent Algorithm (SDA) \cite{Murota2003}}
	\end{algorithm}

The SDA shown in Algorithm 1 is directly devised based on the optimality criterion \eqref{eq:OptCri}. It starts with $\rv^{(0)}$, an arbitrary point in $\Ru$, and moves along the descent direction in each iteration. The authors in \cite{Milo2012} also studied the fairness in CDE problem, where algorithm DA as shown in Algorithm 2 was proposed for solving problem~\eqref{eq:obj}. SDA differs from DA in the following aspects:

\begin{enumerate}[(a)]
    \item Running SDA requires the value of $\alpha$ and a starting point $\rv^{(0)}$ in $\Ru$; Running DA only requires the value of $\alpha$.
    \item For SDA, $\rv^{(k)}\in \Ru$; For DA, $\rv^{(k)}\in P(g_\alpha)$. Since $\overline{\Ru}=B(g_\alpha)$, SDA searches the minimizer in the submodular base polyhedron, while DA searches the minimizer in the submodular polyhedron.
    \item The number of iterations in SDA is bounded by $\frac{\|\rv^{(0)}-\rv^{*}\|_1}{2}$\cite{Murota2003}; The number of iterations in DA is exactly $\alpha$.
\end{enumerate}

	\begin{algorithm} [t]
	\label{algo:DA}
	\small
	\SetAlgoLined
	\SetKwFor{For}{for}{do}{endfor}
	\BlankLine
        $\rv^{(0)}=\Zero$\;
        \For {$k=1$ \emph{\KwTo} $\alpha$} {
            Find $j^*$ such that
                \begin{equation}
                    j^*\in\arg\min\{f_j(r_j+1)-f_j(r_j) \colon j\in\C, \rv+\ev_j\in P(g_4) \};  \nonumber
                \end{equation}
            $\rv^{(k+1)}=\rv^{(k)}+\ev_{j^*}$\;
        }
	
	\caption{Deterministic Algorithm (DA) \cite{Milo2012}}
	\end{algorithm}

Aspect (a) implies that if only $\alpha$ is known DA can be run independently while SDA requires the running of another algorithm to find $\rv^{(0)}\in\Ru$. However, finding $\rv^{(0)}$ is not difficult. It can be accomplished by the randomized algorithm proposed in \cite{Ozgul2011}. If the problem is to find the fairest solution in the minimum sum-rate strategies, i.e., $\alpha=\hat{\alpha}$, there are many existing methods that find the value of $\hat{\alpha}$, and most of them returns $\hat{\alpha}$ with a strategy $\rv^{(0)}\in\Rum$, e.g., the divide-and-conquer algorithm in \cite{MiloDivConq2011}, the randomized algorithms in \cite{SprintRand2010,AbediniNonMinRank2012}. It should be also noted that if SDA and DA are applied after obtaining $\alpha$ and the initial point $\rv^{(0)}\in\Ru$ by the methods in \cite{Ozgul2011,MiloDivConq2011,SprintRand2010,AbediniNonMinRank2012} the knowledge of $\rv^{(0)}$ will be discarded in DA. Aspect (c) is in fact a result of aspect (b). It implies that SDA converges faster than DA if
\begin{equation}
L_1(\alpha)=\max\{\|\xv-\yv\|_1 \colon \xv,\yv\in\Ru\},
\end{equation}
the $l_1$-size of $\Ru$, is smaller than $\alpha$.

In SDA, checking whether $\rv^{(k)}-\ev_u+\ev_v\in\Ru$ is equivalent to checking whether $\rv^{(k)}-\ev_u+\ev_v\in P(g_\alpha)$. Therefore, we can use the same feasibility checking algorithm as in \cite{Milo2012}. The method is to run a submodular function minimization (SFM) algorithm.\footnote{In both \cite{Court2011,Milo2012}, it was shown that checking whether $\rv\in\Ru$ was equivalent to a submodular function minimization problem. This paper applies the same feasibility checking algorithm as in \cite{Court2011,Milo2012}. For more details, we refer the reader to Algorithms IV.3 and IV.4 in \cite{Court2011}. }

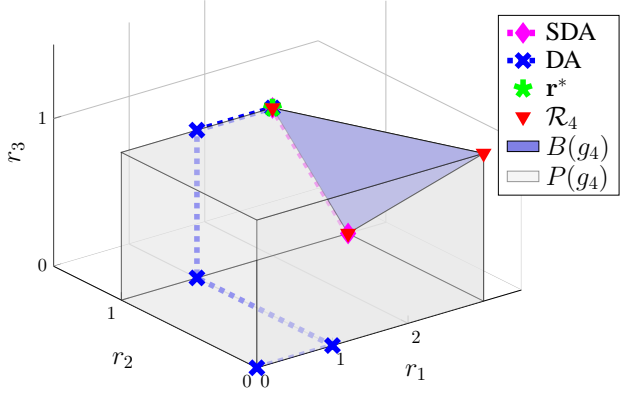
\begin{figure}[tbp]
	\centering
    \scalebox{0.7}{
%
%
%
\definecolor{mycolor1}{rgb}{1,0,1}%
\definecolor{mycolor2}{rgb}{0.5,0.5,0.9}%
\begin{tikzpicture}

\begin{axis}[%
width=3.5in,
height=3in,
view={322.5}{30},
scale only axis,
xmin=0,
xmax=3.5,
xlabel={\Large $r_1$},
xtick={0, 1, 2},
xmajorgrids,
ymin=0,
ymax=1.5,
ylabel={\Large $r_2$},
ytick={0, 1, 2},
ymajorgrids,
zmin=0,
zmax=1.5,
zlabel={\Large $r_3$},
ztick={0, 1, 2},
zmajorgrids,
axis x line*=bottom,
axis y line*=left,
axis z line*=left,
legend style={at={(0.95,0.88)},anchor=north west,draw=black,fill=white,legend cell align=left}
]

\addplot3 [
color=mycolor1,
dashed,
line width=3.5pt,
mark size=3pt,
mark=diamond,
mark options={solid}]
table[row sep=crcr] {
3 1 0\\
2 1 1\\
};
\addlegendentry{\Large SDA};

\addplot3 [
color=blue,
dashed,
line width=3.0pt,
mark size=5.0pt,
mark=x,
mark options={solid}]
table[row sep=crcr] {
0 0 0\\
1 0 0\\
1 1 0\\
1 1 1\\
2 1 1\\
};
\addlegendentry{\Large DA};

\addplot3 [
color=green,
line width=3.0pt,
mark size=5.0pt,
only marks,
mark=star,
mark options={solid}]
table[row sep=crcr] {
2 1 1\\
};
\addlegendentry{\Large $\rv^*$};

\addplot3 [
color=red,
line width=3.0pt,
only marks,
mark=triangle,
mark options={solid,,rotate=180}]
table[row sep=crcr] {
2 1 1\\
3 0 1\\
3 1 0\\
};
\addlegendentry{\Large $\R_4$};

\addplot3[area legend,solid,fill=mycolor2,draw=black]
table[row sep=crcr]{
x y z\\
2 1 1 \\
3 0 1 \\
3 1 0 \\
};

\addlegendentry{\Large $B(g_4)$};

\addplot3[area legend,solid,fill=white!90!black,opacity=4.000000e-01,draw=black]
table[row sep=crcr]{
x y z\\
0 0 0 \\
3 0 0 \\
3 1 0 \\
0 1 0 \\
0 0 0 \\
};

\addlegendentry{\Large $P(g_4)$};

\addplot3[solid,fill=white!90!black,opacity=4.000000e-01,draw=black,forget plot]
table[row sep=crcr]{
x y z\\
0 0 0 \\
0 1 0 \\
0 1 1 \\
0 0 1 \\
0 0 0 \\
};

\addplot3[solid,fill=white!90!black,opacity=4.000000e-01,draw=black,forget plot]
table[row sep=crcr]{
x y z\\
0 0 0 \\
3 0 0 \\
3 0 1 \\
0 0 1 \\
0 0 0 \\
};

\addplot3[solid,fill=white!90!black,opacity=4.000000e-01,draw=black,forget plot]
table[row sep=crcr]{
x y z\\
0 0 0 \\
3 0 0 \\
3 0 1 \\
0 0 1 \\
0 0 0 \\
};

\addplot3[solid,fill=white!90!black,opacity=4.000000e-01,draw=black,forget plot]
table[row sep=crcr]{
x y z\\
3 0 0 \\
3 1 0 \\
3 0 1 \\
3 0 0 \\
};

\addplot3[solid,fill=white!90!black,opacity=4.000000e-01,draw=black,forget plot]
table[row sep=crcr]{
x y z\\
0 1 0 \\
0 1 1 \\
2 1 1 \\
3 1 0 \\
0 1 0 \\
};

\addplot3[solid,fill=white!90!black,opacity=5.000000e-01,draw=black,forget plot]
table[row sep=crcr]{
x y z\\
0 0 1 \\
3 0 1 \\
2 1 1 \\
0 1 1 \\
0 0 1 \\
};

\end{axis}
\end{tikzpicture}%
}
	\caption{The searching paths of SDA and DA for problem~\eqref{eq:obj} with $\alpha=4$ in the CDE system in Fig.~\ref{fig:CDESystem}. The number of iterations is $1$ for SDA and $4$ for DA. For SDA, $\rv^{(k)}\in B(g_4)\cap\Z^K$, i.e., $\rv^{(k)}\in\R_4$; For DA, $\rv^{(k)}\in P(g_4)\cap\Z^K$.}
	\label{fig:Paths}
\end{figure}

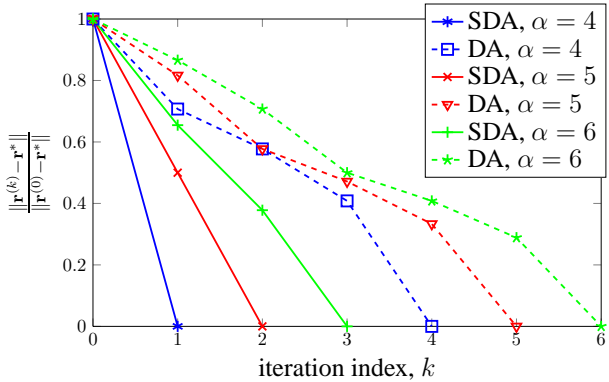
\begin{figure}[tbp]
	\centering
    \scalebox{0.7}{
%
%
\begin{tikzpicture}

\begin{axis}[%
width=3.8in,
height=2.3in,
scale only axis,
xmin=0,
xmax=6,
xlabel={\Large iteration index, $k$},
xtick={0, 1, 2, 3, 4, 5, 6},
ymin=0,
ymax=1,
ylabel={\Large $\frac{\|\rv^{(k)}-\rv^*\|}{\|\rv^{(0)}-\rv^*\|}$},
legend style={at={(1.01,1.05)},draw=black,fill=white,legend cell align=left}
]
\addplot [
color=blue,
solid,
line width=1.0pt,
mark=asterisk,
mark size=3.0pt,
mark options={solid}
]
table[row sep=crcr]{
0 1\\
1 0\\
};
\addlegendentry{\Large SDA, $\alpha=4$};

\addplot [
color=blue,
dashed,
line width=1.0pt,
mark=square,
mark size=3.0pt,
mark options={solid}
]
table[row sep=crcr]{
0 1\\
1 0.707106781186548\\
2 0.577350269189626\\
3 0.408248290463863\\
4 0\\
};
\addlegendentry{\Large DA, $\alpha=4$};

\addplot [
color=red,
solid,
line width=1.0pt,
mark=x,
mark size=3.0pt,
mark options={solid}
]
table[row sep=crcr]{
0 1\\
1 0.5\\
2 0\\
};
\addlegendentry{\Large SDA, $\alpha=5$};

\addplot [
color=red,
dashed,
line width=1.0pt,
mark=triangle,
mark size=3.0pt,
mark options={solid,,rotate=180}
]
table[row sep=crcr]{
0 1\\
1 0.816496580927726\\
2 0.577350269189626\\
3 0.471404520791032\\
4 0.333333333333333\\
5 0\\
};
\addlegendentry{\Large DA, $\alpha=5$};

\addplot [
color=green,
solid,
line width=1.0pt,
mark=+,
mark size=3.0pt,
mark options={solid}
]
table[row sep=crcr]{
0 1\\
1 0.654653670707977\\
2 0.377964473009227\\
3 0\\
};
\addlegendentry{\Large SDA, $\alpha=6$};

\addplot [
color=green,
dashed,
line width=1.0pt,
mark=star,
mark size=3.0pt,
mark options={solid}
]
table[row sep=crcr]{
0 1\\
1 0.866025403784439\\
2 0.707106781186547\\
3 0.5\\
4 0.408248290463863\\
5 0.288675134594813\\
6 0\\
};
\addlegendentry{\Large DA, $\alpha=6$};

\end{axis}
\end{tikzpicture}%
}
	\caption{Convergence performance of SDA and DA for problem~\eqref{eq:obj} in the CDE system in Fig.~\ref{fig:CDESystem}: $\frac{\|\rv^{(k)}-\rv^*\|}{\|\rv^{(0)}-\rv^*\|}$, the normalized error, when $\alpha=4$, $5$ and $6$. It can be seen that the number of iterations in DA equals to the value of $\alpha$. The number of iterations in SDA depends on the distance between the starting point $\rv^{(0)}$ and minimizer $\rv^*$. It is less than DA for all three values of $\alpha$. }
	\label{fig:ErrSamp}
\end{figure}

\begin{example} \label{ex:Converge}
Consider problem~\eqref{eq:obj} when $\alpha=4$ for the CDE system in Fig.~\ref{fig:CDESystem}. By applying SDA algorithm, we get the estimation sequence $\{\rv^{(k)}\}=\{(3,1,0),(2,1,1)\}$, where the starting point $\rv^{(0)}=(3,1,0)$ is obtained by running the algorithm in \cite{Ozgul2011}; By applying DA algorithm, we get the estimation sequence $\{\rv^{(k)}\}=\{(0,0,0),(1,0,0),(1,1,0),(1,1,1),(2,1,1)\}$. It can be seen that SDA converges faster than DA. We plot the searching paths indicated by sequence $\{\rv^{(k)}\}$ for both SDA and DA in Fig.~\ref{fig:Paths}. It clearly shows that SDA searches the minimizers in the submodular base polyhedron $B(g_4)$, while DA searches the minimizer in the submodular polyhedron $P(g_4)$. We then run SDA and DA for $\alpha=5$ and $\alpha=6$. We plot $\frac{\|\rv^{(k)}-\rv^*\|}{\|\rv^{(0)}-\rv^*\|}$, the normalized errors, in Fig.~\ref{fig:ErrSamp}. It shows that SDA requires less number of iterations than DA before reaching the minimizer $\rv^*$.
\end{example}

Although SDA converges faster than DA, the complexity of SDA could be higher than that of DA. Let $\zeta$ denote the complexity of running SFM algorithm for the feasibility check. We assume that $\zeta$ includes the complexity of the measurement of the objective function if the results of the feasibility check is true. For SDA, the complexity is bounded by $O(K^2\cdot\zeta\cdot L_1(\alpha))$ \cite{Murota2003}. Note, $O(K^2\cdot\zeta\cdot L_1(\alpha))$ is the maximum complexity of SDA for two reasons: one is that $L_1(\alpha)$ is the maximum $l_1$-norm of $\Ru$; The other is that we only need to run SFM algorithm (to check the feasibility) for those $(u,v)$ such that $\rv^{(k)}-\ev_u+\ev_v\geq{\Zero}$. For DA, the complexity is exactly $O(K\cdot\zeta\cdot \alpha)$. Therefore, the main difference in complexity between SDA and DA is the number of runs of SFM algorithm. Since $L_1(\alpha)$ grows with $\alpha$, the computation load of SDA may be heavier than that of DA when $L_1(\alpha)$ is comparable to or larger than $\alpha$.

Consider problem~\eqref{eq:obj} when $\alpha=\hat{\alpha}$, i.e., the problem of finding the fairest solution in the minimum sum-rate strategies. In this case, $L_1(\hat{\alpha})$ is minimum, and $\hat{\alpha}$ is proportional to $N$, the number of packets \cite{Roua2010}. Therefore, the computation loads of SDA and DA grow with both $K$ and $N$, and the complexity growth in SDA is larger than DA.

\begin{figure}[t]
	\centering
        \subfigure[Convergence performance of SDA and DA: the average number of iterations before reaching the minimizer $\rv^*$]{\includegraphics[height=5.2cm,width=8.3cm]{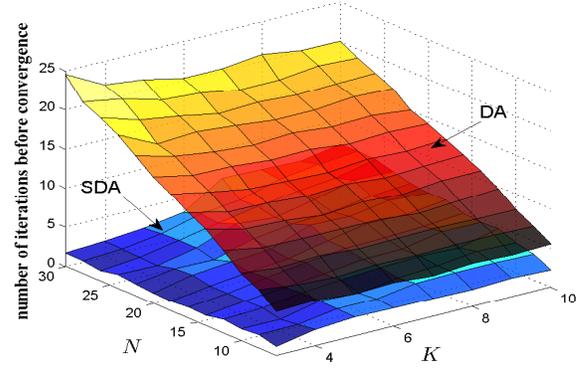}} 
        \subfigure[Time complexity of SDA and DA: the average number of runs of SFM algorithm]{\includegraphics[height=5.2cm,width=8.3cm]{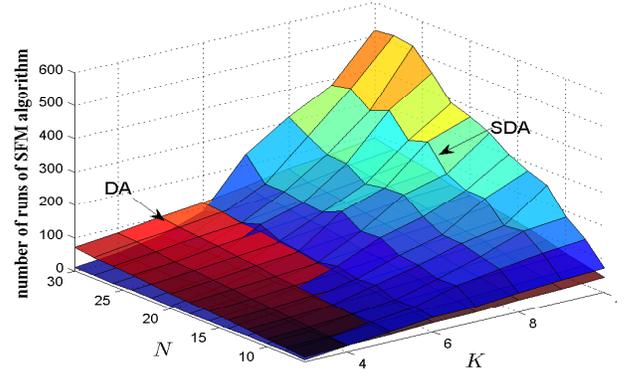}}
	\caption{Comparisons of SDA and DA when applied to problem~\eqref{eq:obj} with $\alpha=\hat{\alpha}$, the problem of finding the fairest solution in the minimum sum-rate strategies. $K$, the number of clients is varying from $3$ to $10$, and $N$, the number of packets, is varying from $6$ to $30$.}
	\label{fig:ITTCComp}
\end{figure}

\begin{example}
We run an experiment to compare SDA and DA in terms of convergence performance and complexity when they are applied to search the fairest solution in the minimum sum-rate strategies. We vary $K$, the number of clients, from $3$ to $10$ and $N$, the number of packets, from $6$ to $30$. For each combination of $K$ and $N$, we repeat the following steps by $20$ times.
\begin{itemize}
    \item Randomly generate the has-set $\Has_j$ for each client subject to the condition $\cup_{j\in\C}\Has_j=\Pak$.
    \item Run the randomized algorithm in \cite{AbediniNonMinRank2012} to obtain the minimum sum-rate $\hat{\alpha}$ and the starting point $\rv^{(0)}\in\Rum$.
    \item Apply SDA and DA to search the fairest transmission strategy $\rv^*$.\footnote{The randomized algorithm in \cite{AbediniNonMinRank2012} returns $\hat{\alpha}$ with a strategy $\rv^{(0)}\in\Rum$. In SDA, both $\hat{\alpha}$ and $\rv^{(0)}$ are used. But, in DA, only $\hat{\alpha}$ is used.}
\end{itemize}
The number of iterations and complexity averaged over repetitions are recorded and shown in Fig.~\ref{fig:ITTCComp}. In Fig.~\ref{fig:ITTCComp}(a), it can be seen that SDA always converges faster than DA and there is a clear growth in the number of iterations in DA with increasing $K$.\footnote{It also implies that $\hat{\alpha}$ grows with $K$ more drastically than $L_1(\hat{\alpha})$.} According to Fig.~\ref{fig:ITTCComp}(b), the complexity of SDA is lower than or close to that of DA when $K$ is lower than $6$. It can be seen that the complexity of SDA is much higher than that of DA when $K$ and $N$ are large. The results in Fig.~\ref{fig:ITTCComp} shows that SDA is more efficient than DA when the number of clients is up to $5$.
\end{example}

\section{Conclusion and Future Work}
We formulated a discrete minimization problem for finding the fairest solution in the constant sum-rate strategies that achieved universal recovery in CDE system. We proved the $M$-convexity of this problem and presented an SDA algorithm for searching the minimizer. We discussed the relationship between our work, the results in \cite{Milo2012} and resource allocation problems with submodular constraints. By a comparison in convergence performance and complexity between SDA and DA, we showed that SDA was more efficient than DA when it applied to the problem of finding the fairest solution in the minimum sum-rate strategies among a small number of clients.

As part of the conclusion, we briefly discuss how the results in this paper can guide the research work on CDE problems in the future. One can study the $M$-convexity of other optimization problems over the feasible region $\Ru$. For example, the weighted sum transmission cost minimization problem $\min\{\mathbf{w}^\intercal\rv\colon\rv\in\Ru\}$, where $\mathbf{w}$ is a weight vector. Since $\mathbf{w}^\intercal\rv$ is separable convex, the $M$-convexity of this problem is directly proved by Theorems~\ref{theo:Mconset} and \ref{theo:Mconfun} in Section~\ref{sec:Main}.\footnote{$\Fu(\rv)$ and $\mathbf{w}^\intercal\rv$ are both separable convex in $\rv$. But, separable convex function is just a special case of $M$-convex function. For the problem where the objective function is non-separable, it still could be $M$-convex as long as the condition in Definition~\ref{def:Mconfun} is satisfied. It should be also noted that SDA can apply to non-separable $M$-convex minimization problems while DA can not.} For solving $M$-convex optimization problem, one can consider algorithms other than SDA, e.g., the algorithms proposed in \cite{Shioura2004,Tamura2005,Moriguchi2011}.\footnote{Some algorithms in \cite{Shioura2004,Tamura2005,Moriguchi2011} are based on the properties other than the optimality criterion, e.g., algorithm in \cite{Moriguchi2011} is devised by utilizing the proximity theorem of $M$-convexity. The works in \cite{Moriguchi2011} also show how to obtain the starting point $\rv^{(0)}$ by constructing a convex extension of the $M$-convex objective function.} The complexity in these algorithms could be lower than SDA, i.e., they could be more efficient than both SDA and DA.

\appendices

\section{}
\label{app:lemma}
\begin{lemma} \label{lemma:app}
For $\Has_j\subset\Pak$ and $\X,\Y\subset\C$ such that $\X\cap\Y\neq\emptyset$ and $\X\cup\Y\neq\C$,
\begin{equation}
 \Big| \bigcap_{j\in\X}\Has_j \Big| + \Big| \bigcap_{j\in\Y}\Has_j \Big|  \leq \Big| \bigcap_{j\in\X\cup\Y}\Has_j \Big| + \Big| \bigcap_{j\in\X\cap\Y}\Has_j \Big| \nonumber
\end{equation}
\end{lemma}
\begin{IEEEproof}
Recall that $\bigcap_{j\in\X}\Has_j\subseteq\bigcap_{j\in\X\cap\Y}\Has_j$ and $\bigcap_{j\in\Y}\Has_j\subseteq\bigcap_{j\in\X\cap\Y}\Has_j$ because $\X\cap\Y\subseteq\X$ and $\X\cap\Y\subseteq\Y$, respectively. We have
\begin{align}
& \quad \Big| (\bigcap_{j\in\X}\Has_j)\cup(\bigcap_{j\in\Y}\Has_j) \Big| \nonumber\\
& \leq \Big| (\bigcap_{j\in\X\cap\Y}\Has_j)\cup(\bigcap_{j\in\X\cap\Y}\Has_j) \Big| \nonumber\\
& = \Big| (\bigcap_{j\in\X\cap\Y}\Has_j) \Big|.
\end{align}
Then, we have
\begin{align}
&\quad \Big| \bigcap_{j\in\X}\Has_j \Big| + \Big| \bigcap_{j\in\Y}\Has_j \Big|   \nonumber\\
&= \Big| (\bigcap_{j\in\X}\Has_j)\cap(\bigcap_{j\in\Y}\Has_j) \Big| + \Big| (\bigcap_{j\in\X}\Has_j)\cup(\bigcap_{j\in\Y}\Has_j) \Big|  \nonumber\\
&\leq \Big| \bigcap_{j\in\X\cup\Y}\Has_j \Big| + \Big| (\bigcap_{j\in\X\cap\Y}\Has_j) \Big|,
\end{align}
which proves Lemma~\ref{lemma:app}.
\end{IEEEproof}

\begin{corollary} \label{coro:app}
For $\Has_j\subset\Pak$ and $\X,\Y\subset\C$ such that $\X\cap\Y\neq\emptyset$ and $\X\cup\Y\neq\C$,
\begin{align}
&\quad \Big| \bigcap_{j\in\C\setminus\X}\Has_j^{c} \Big| + \Big| \bigcap_{j\in\C\setminus\Y}\Has_j^{c} \Big| \nonumber \\
&\leq \Big| \bigcap_{j\in(\C\setminus\X)\cup(\C\setminus\Y)}\Has_j^{c} \Big| + \Big| \bigcap_{j\in(\C\setminus\X)\cap(\C\setminus\Y)}\Has_j^{c} \Big| \nonumber
\end{align}
\end{corollary}
\begin{IEEEproof}
The proof can be show by substituting $\X$ by $\C\setminus\X$, $\Y$ by $\C\setminus\Y$ and $\Has_j$ by $\Has_j^c$ in Lemma~\ref{lemma:app}.
\end{IEEEproof}

\bibliography{IEEEabrv,CDEFairBIB}


\end{document}